\documentclass[aps,pre,amsmath,amssymb,twocolumn,superscriptaddress,a4paper,showpacs,floatfix]{revtex4-1}

\usepackage[utf8]{inputenc}

\usepackage{microtype}

\usepackage{amsfonts}
\usepackage{amsmath}
\usepackage{amssymb}

\usepackage{graphicx}% Include figure files
\usepackage{dcolumn} % Align table columns on decimal point
\usepackage{hyperref}% add hypertext capabilities
\usepackage{natbib}

\usepackage{blindtext}

\newcommand{\vect}[1]{{\mathbf{#1}}}
\newcommand{\ind}[1]{{\mathrm{#1}}}

\begin{document}

\title{Particle and particle pair dispersion in turbulence modeled with spatially and temporally correlated stochastic processes}

\author{Thomas Burgener}
\email{buthomas@ethz.ch}
\affiliation{Computational Physics, IfB, ETH-H\"onggerberg, Schafmattstrasse 6, 8093 Z\"urich, Switzerland}
\author{Dirk Kadau}
\email{dkadau@ethz.ch}
\affiliation{Computational Physics, IfB, ETH-H\"onggerberg, Schafmattstrasse 6, 8093 Z\"urich, Switzerland}
\author{Hans J. Herrmann}
\email{hans@ifb.baug.ethz.ch}
\affiliation{Computational Physics, IfB, ETH-H\"onggerberg, Schafmattstrasse 6, 8093 Z\"urich, Switzerland}
\affiliation{Departamento de Fis\'ica, Universidade Federal do Cear\'a, Campus do Pici, 60451-970 Fortaleza, Cear\'a, Brazil}

\date{\today}

\begin{abstract}
  In this paper we present a new model for modeling the diffusion and relative dispersion of particles in homogeneous isotropic turbulence. We use an Heisenberg-like Hamiltonian to incorporate spatial correlations between fluid particles, which are modeled by stochastic processes correlated in time. We are able to reproduce the ballistic regime in the mean squared displacement of single particles and the transition to a normal diffusion regime for long times. For the dispersion of particle pairs we find a $t^{2}$-dependence of the mean squared separation at short times and a $t$-dependence for long ones. For intermediate times indications for a Richardson $t^{3}$ law are observed in certain situations. Finally the influence of inertia of real particles on the dispersion is investigated.
\end{abstract}

% 47.27.T-	Turbulent transport processes
% 47.27.E-	Turbulence simulation and modeling
% 83.10.Rs	Computer simulation of molecular and particle dynamics
% 45.70.-n	Granular systems

%\keywords{}

\maketitle

\section{Introduction}
Relative dispersion of particles in turbulent flows is of key importance in a variety of natural and industrial processes, ranging from the spreading of clouds in the atmosphere or pollutants in the ocean to mixing in pneumatic conveyors or production of nanoparticles in flames. Therefore the diffusion of particles and the relative dispersion of pairs of particles in homogeneous isotropic turbulence have been investigated intensively \cite{Richardson1926,Obukhov1941,Batchelor1950,Squires1991a,Sawford2001,Sawford2008,Salazar2009}. Although considerable progress in understanding these processes has been made \cite{Sawford2001,Salazar2009}, we still lack the fundamental understanding of the observed phenomena.

Recent advances in direct numerical simulations (DNS) of isotropic turbulence \cite{Yeung2006} made it possible to investigate turbulent dispersion \cite{Sawford2008} at increasingly high Reynolds numbers. But because of their extremely high demand of computer resources, DNS calculations are unfortunately not applicable in most situations. To circumvent this difficulty many different techniques for turbulence modeling \cite{Pope2000} have been invented.

In this paper we present a new model, which is especially useful for the investigation of particle dispersion at high Reynolds numbers. Based on a system of stochastic differential equations (SDEs) introduced by A.~M. Reynolds \cite{Reynolds2003a,Sawford1991,Pope1990} that describes the fluid by a set of tracer particles, we use a Heisenberg-like Hamiltonian to incorporate spatial correlations between fluid particles. This enables us to generate a three dimensional turbulent velocity field for which the spatial structure functions show the desired behaviour. We then investigate the dispersion of single fluid particles by measuring the mean squared displacement (MSD) as well as the relative dispersion of pairs of fluid particles through their mean squared separation (MSS). Finally we replace the fluid particles by real heavy particles and investigate the influence of the inertia of these particles on the MSD as well as the MSS.

Turbulent dispersion was first investigated by Taylor \cite{Taylor1922}, who studied the diffusion of single particles in isotropic turbulence. Later Richardson \cite{Richardson1926}, Obukhov \cite{Obukhov1941}, and Batchelor \cite{Batchelor1950} extended the investigation to multiple particles. Many measurements of the spread of clouds or puffs, as well as the separation of rather large tracer particles have been performed (see e.g.\ Ref.~\cite{Monin1975}, pp.~556-567). Recent advances in experimental techniques made it possible to track an increasing number of small tracers at high Reynolds numbers in laboratory flows (see e.g.\ Ref.~\cite{Salazar2009}). In three dimensions experiments are often performed in a tank filled with water and turbulence is generated e.g.\ by counter-rotating baffled disks \cite{Bourgoin2006} or rotating propellers located in the corners of the tank \cite{Berg2006}. In two dimensions experiments have been performed with thin layers of conducting fluids where turbulence is generated through electromagnetic forces \cite{Jullien1999}. Simulations most often use DNS methods in three dimensions, where the Navier-Stokes equations are solved directly and turbulence is generated by a spectral forcing technique \cite{Yeung1994,Sawford2008}. Kinematic simulations \cite{Fung1992} are another technique frequently used in simulations. Most of these publications focus on the investigation of the MSS of particles as well as the pair-separation probability density functions. More detailed introductions can be found e.g.\ in the reviews of Sawford \cite{Sawford2001} and Salazar \& Collins \cite{Salazar2009}.

\section{Spatial correlations between fluid particles}
The spatial and temporal structure of homogeneous and isotropic turbulence is highly non-trivial. Of particular interest are the spatial velocity structure functions. Let $\vect{u}\left( \vect{x},t \right)$ be the fluid velocity at time $t$ and position $\vect{x}$. The (longitudinal) velocity structure functions are then defined as
\begin{equation}
D_{n}\left( r \right)=\Bigl< \bigl| ( \vect{u}(\vect{x}+\vect{r},t) - \vect{u}(\vect{x},t) ) \cdot \vect{r} \bigr|^{n} \Bigr>,
  \label{eqn:Dn_Def}
\end{equation}
where $\vect{r}$ is a vector connecting two points in the fluid field that are separated at a distance $r=\lvert\vect{r}\rvert$. According to the Kolmogorov similarity theory \cite{Monin1975} these structure functions depend only on the mean energy dissipation rate $\langle\varepsilon\rangle$ and the distance $r$, i.e.
\begin{equation}
  D_{n}\left( r \right)=C_{n}\bigl( \langle\varepsilon\rangle\cdot r \bigr)^{\zeta_{n}}.
  \label{eqn:Dn_Kolmogorov}
\end{equation}
Further by Kolmogorov's second hypothesis the exponents $\zeta_{n}$ should be given by
\begin{equation}
  \zeta_{n}=\frac{n}{3}.
  \label{eqn:zeta_n_3}
\end{equation}
However, several measurements \cite{Anselmet1984} show that for $n>3$ the exponents are smaller than the Kolmogorov values. Most commonly studied \cite{Saddoughi1994} is the second order structure function, for which it can be shown \cite{Pope2000} that
\begin{equation}
  D_{2}\left( r \right) = 
  \begin{cases}
    \frac{\langle\varepsilon\rangle}{15\nu}r^{2} & r\ll\eta\\[0.6ex]
    C_{2}\bigl( \langle\varepsilon\rangle\cdot r \bigr)^{2/3} & \eta\ll r\ll L\\[0.9ex]
    2\sigma_{\ind{u}}^{2} & L\ll r
  \end{cases},
  \label{eqn:D2_cases}
\end{equation}
where $\nu$ is the kinematic fluid viscosity, $\sigma_{\ind{u}}^{2}$ the variance of the fluid velocity, $\eta=(\nu^{3}/\langle\varepsilon\rangle)^{1/4}$ the Kolmogorov length scale and $L=\sigma_{\ind{u}}^{3}/\langle\varepsilon\rangle$ the turbulence length scale. The value of the Kolmogorov constant \cite{Pope2000} is $C_{2}=2.0$.

To generate a turbulent velocity field we start with a model introduced by Reynolds \cite{Reynolds2003a}. This model describes well the measured \cite{LaPorta2001} velocity and acceleration distributions of Lagrangian tracer particles in fully developed turbulence by modeling the motion of non-interacting fluid tracer particles that are self-correlated in time. It consists of a set of stochastic differential equations (SDEs) for the logarithm of the local energy dissipation rate $\chi=\ln \left( \varepsilon/\langle\varepsilon\rangle \right)$ as well as equations for one component of the acceleration, velocity and position of a fluid particle:
\begin{subequations}
  \label{eqn:sde}
  \begin{gather}
    d\chi = -\left( \chi - \left< \chi \right> \right)T_{\chi}^{-1}dt+\sqrt{2\sigma_{\chi}^{2}T_{\chi}^{-1}}d\xi_1\label{eqn:sde_chi}\\
    da_{\ind{t}} = -\left( T_{L}^{-1} + t_{\eta}^{-1} -\sigma_{a_{\ind{t}}|\varepsilon}^{-1}\frac{d\sigma_{a_{\ind{t}}|\varepsilon}}{dt} \right) a_{\ind{t}} dt -T_{L}^{-1} t_{\eta}^{-1} u_{\ind{t}} dt\nonumber\\*
    + \sqrt{2\sigma_{\ind{u}}^2\left( T_{L}^{-1} + t_{\eta}^{-1} \right)T_{L}^{-1} t_{\eta}^{-1} }d\xi_2\label{eqn:sde_a}\\
    du_{\ind{t}} = a_{\ind{t}}dt\label{eqn:sde_u}\\
    dx_{\ind{t}} = u_{\ind{t}}dt.\label{eqn:sde_x}
  \end{gather}
\end{subequations}
The variance of $\chi$ in Eq.~\eqref{eqn:sde_chi} is approximated \cite{Yeung1989} by $\sigma_{\chi}^2=-0.354+0.289\log R_{\lambda}$, where the Reynolds number at the Taylor microscale is given by $R_{\lambda}=\sqrt{15\sigma_{\ind{u}}L/\langle\varepsilon\rangle}$. The mean value is given by $\left< \chi \right> = -0.5 \sigma_{\chi}^2$, and the relaxation time scale by $T_{\chi}=2\sigma_{u}^2/(C_0\left< \varepsilon \right>)$. Further in Eq.~\eqref{eqn:sde_a} the energy-containing time scale (sometime also called integral time scale) is given by $T_{\ind{L}}=2\sigma_{\ind{u}}^{2}/(C_{0}\varepsilon)$, the energy-dissipation time scale by $t_{\eta}=C_{0}\nu^{1/2}/(2a_{0}\varepsilon^{1/2})$, and the conditional acceleration variance is $\sigma_{a_{\ind{t}}|\varepsilon}^{2}=a_0\varepsilon^{3/2}\nu^{1/2}$. The two universal Lagrangian velocity structure constants are given as $a_{0}=3.3$ and $C_{0}=7.0$. Finally $d\xi_{1}$ and $d\xi_{2}$ are two independent Wiener processes, i.e. Gaussian distributed random numbers with zero mean and variance $dt$.

In Ref.~\cite{Burgener2011} we already used this model to investigate the mixing of heavy particles in a turbulent channel flow due to intrinsic fluctuations in the fluid velocity. There we ``glued'' one fluid tracer particle to every real particle and used the resulting velocity vector $\vect{u}_{\ind{t}}$ in an empirical drag law to get a force acting on the real particle. The equations of motion of the fluid particles were integrated according to Eqs.~\eqref{eqn:sde_chi}--\eqref{eqn:sde_u} and their positions were determined by the positions of the real particles. The biggest deficiency of this model is that there are no spatial correlations between the fluid particles. Calculating the velocity structure functions \eqref{eqn:Dn_Def} in this model results in
\begin{equation}
  D_{n}\left( r \right)=\frac{2^{n}\,\Gamma\left( \frac{n+1}{2} \right)}{\sqrt{\pi}}\cdot\sigma_{\ind{u}}^{n}=\text{const.},
  \label{eqn:Dn_Gaussian}
\end{equation}
where $\Gamma\left( x \right)$ is the Gamma function, which are equal to the central absolute moments of a Gaussian distribution $\mathcal{N}\left( 0,2\sigma_{\ind{u}}^{2} \right)$, because the components of $\vect{u}_{\ind{t}}$ are (by construction) Gaussian distributed.

Our basic idea for introducing spatial correlation between the fluid particles is to formulate an Heisenberg-like Hamiltonian which is then minimized. In more detail let us consider a set of $n_{\ind{p}}$ fluid particles at positions $\vect{x}_{i}$ in a cubic volume of linear size $L_{\ind{v}}$ with periodic boundary conditions. Every fluid particle $i$ has a velocity $\vect{u}_{i}$, acceleration $\vect{a}_{i}$ and (local) energy dissipation $\varepsilon_{i}$. These particles are independently integrated in time according to Eqs.~\eqref{eqn:sde}. We then define an Hamiltonian
\begin{equation}
  \mathcal{H}=-\sum_{i\neq j}J\left( r_{ij} \right) \frac{\vect{u}_{i}\cdot\vect{u}_{j}}{\lvert\vect{u}_{i}\rvert\,\lvert\vect{u}_{j}\rvert},
  \label{eqn:Hamiltonian}
\end{equation}
where $r_{ij}=\lvert\vect{x}_{j}-\vect{x}_{i}\rvert$ is the distance between particle $i$ and $j$, and $J\left( r_{ij} \right)$ is a distance dependent coupling function. The choice of this coupling function is not known a priori and heavily influences the resulting correlations. The minimization of the Hamiltonian $\mathcal{H}$ is achieved by a standard Metropolis algorithm \cite{Metropolis1953}: A pair of fluid particles is randomly selected and a new ``configuration'' is proposed consisting of interchanging the two particles. Then the change in energy $\Delta E$ is calculated and the interchange is accepted with probability
\begin{equation}
  p\left( \Delta E \right) = \min\Bigl( 1,\exp(-\Delta E / T) \Bigr),
  \label{eqn:AcceptanceProbability}
\end{equation}
where $T$ is an artificial ``temperature'', chosen to be ``low''. This kind of correlation steps basically just aligns the fluid particles to minimize the Hamiltonian \eqref{eqn:Hamiltonian}. This ensures that the total kinetic energy and momentum of the system are conserved, and additionally also the temporal statistics of the SDEs are (on average) unchanged. The correlation steps are then included in the whole algorithm, such that between two time integration steps of all the SDEs a certain number of correlations steps is performed. The number of correlation steps is chosen such that after one Kolmogorov time $\tau_{\eta}=(\nu/\langle\varepsilon\rangle)^{1/2}$ about $n_{\ind{p}}$ correlation steps are performed.

The coupling function $J\left( r_{ij} \right)$ in Eq.~\eqref{eqn:Hamiltonian} has to be adjusted such that the desired structure functions are obtained. Here $J\left( r_{ij} \right)$ has been chosen as a power law
\begin{equation}
  J\left( r_{ij} \right) = J_{0}\,r_{ij}^{\alpha}.
  \label{eqn:powerLaw}
\end{equation}
Other choices for $J\left( r_{ij} \right)$ were used as well, but we found a power law to give the best results and to be the easiest to control. Changing the parameter $\alpha$ leads to a horizontal shift of the resulting structure functions and variations of $J_{0}$ change the exponents of $D_{n}\left( r \right)$. Since the model Hamiltonian \eqref{eqn:Hamiltonian} is purely empirical, we have no concise explanation why a power law for $J(r_{ij})$ gives the best result. It could be possible that other functional forms for the coupling function may give similar results.

We performed simulations with $n_{\ind{p}}=27000$ fluid particles for three different Reynolds numbers at the Taylor microscale $R_{\lambda}=372$, $740$ and $1115$. The parameters used in these simulations are listed in Tab.~\ref{tab:parameter}. %
\begin{table}
  \caption{\label{tab:parameter}Parameters used throughout the simulations in this paper.}
  \begin{ruledtabular}
    \begin{tabular}{llll}
      $R_{\lambda}$ & 372 & 740 & 1115\\[2ex]
      $\sigma_{\ind{u}}^{2}$      & $0.4802\,\mathrm{m}/\mathrm{s}$       & $0.9604\,\mathrm{m}/\mathrm{s}$       & $1.4406\,\mathrm{m}/\mathrm{s}$      \\
      $\langle\varepsilon\rangle$ & $25.0\,\mathrm{m}^{2}/\mathrm{s}^{3}$ & $25.0\,\mathrm{m}^{2}/\mathrm{s}^{3}$ & $25.0\,\mathrm{m}^{2}/\mathrm{s}^{3}$\\
      $\nu$                       & $0.000001\,\mathrm{m}^{2}/\mathrm{s}$ & $0.000001\,\mathrm{m}^{2}/\mathrm{s}$ & $0.000001\,\mathrm{m}^{2}/\mathrm{s}$\\[2ex]
      $J_{0}$                     & $0.00001$                             & $0.00001$                             & $0.00005$                            \\
      $\alpha$                    & $-1.61$                               & $-2.35$                               & $-3.31$                              \\
      $r_{\ind{c}}$               & $0.01\,\mathrm{m}$                    & $0.02\,\mathrm{m}$                    & $0.03\,\mathrm{m}$                   \\[2ex]
      $L_{\ind{v}}$               & $1.0\,\mathrm{m}$                     & $1.0\,\mathrm{m}$                     & $1.0\,\mathrm{m}$                    \\
      $n_{\ind{p}}$               & $27000$                               & $27000$                               & $27000$                              \\
      $T$                         & $0.01$                                & $0.01$                                & $0.01$                               \\
    \end{tabular}
  \end{ruledtabular}
\end{table}
The parameter $r_{\ind{c}}$ is a cut-off distance for the power law \eqref{eqn:powerLaw}. One reason for such a cut-off is to reduce the complexity of the simulation. Calculating the energy in the Hamiltonian \eqref{eqn:Hamiltonian} would grow quadratically with the number of particles if $J\left( r_{ij} \right)$ had an infinite range. Introducing a cut-off reduces the complexity of this calculation drastically, since only particle pairs within a distance $r_{\ind{c}}$ need to be considered. The second reason for using a cut-off is that the infinite range of the power law gives rise to too strong correlations between the fluid particles at large distances. We found that choosing a cut-off distance $r_{\ind{c}}$, which is a bit smaller than the turbulence length scale $L$ to give the best results. Since $L$ increases with increasing $R_{\ind{\lambda}}$, the values of $r_{\ind{c}}$ are different for the two Reynolds numbers.

\begin{figure}
  \includegraphics{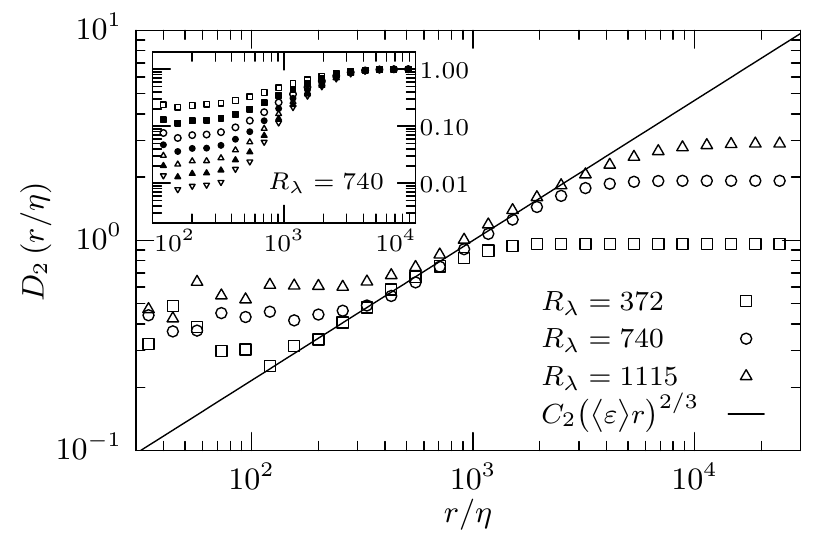}
  \caption{Measured second order spatial velocity structure function $D_{2}\left( r \right)$ for Reynolds numbers $R_{\lambda}=372$, $740$ and $1115$. For intermediate distances the desired $r^{2/3}$ scaling is observed. For smaller distances the structure function does not show the behaviour given in Eq.~\eqref{eqn:D2_cases}, because our model is not able to correlate the velocities sufficiently at small distances. The insets show the higher order velocity structure functions rescaled with the large distance limits from Eq.~\eqref{eqn:Dn_Gaussian} for $n=2,\dots,8$ (top to bottom) for $R_{\lambda}=740$. The increase of the exponents of the power laws is clearly visible.}
  \label{fig:D2_R}
\end{figure}

\begin{figure}
  \includegraphics{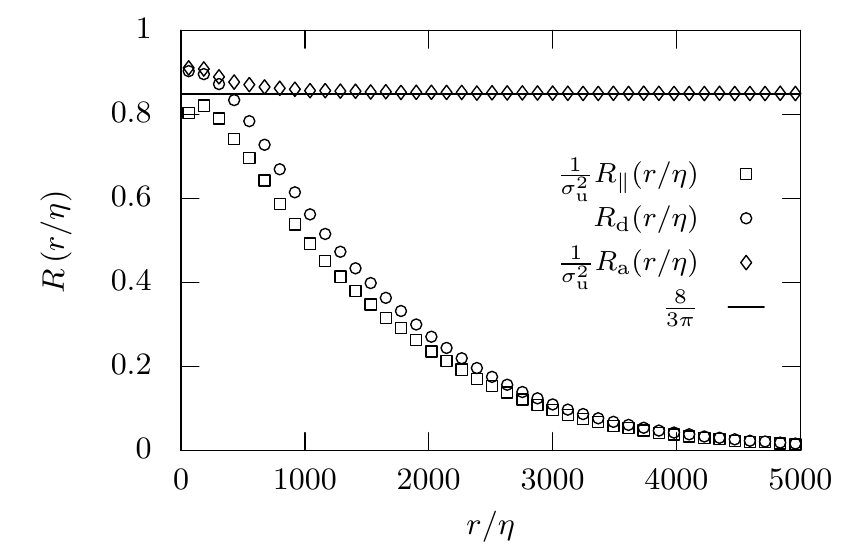}
  \caption{Measured spatial velocity correlation functions at $R_{\lambda}=740$. The velocity correlation $R_{\parallel}\left( r/\eta \right)$ is too small at small distances. First, this is due to the facts that our model is not able to perfectly correlate the directions of the fluid particle velocities for small $r$ as shown by $R_{\ind{d}}\left( r/\eta \right)$. Second, the Hamiltonian \eqref{eqn:Hamiltonian} does not correlate the magnitudes of the fluid velocities which leads to $R_{\ind{a}}\left( r/\eta \right)$ being almost independent of $r$.}
  \label{fig:R_740}
\end{figure}

Fig.~\ref{fig:D2_R} shows the second order structure function $D_{2}\left( r/\eta \right)$ for the three considered Reynolds numbers $R_{\lambda}=372$, $740$ and $1115$.
Depending on this Reynolds number, we can observe the desired dependence for large distances given by Eq.~\eqref{eqn:D2_cases}. Unfortunately for small distances the measured structure functions do not agree with the desired functions \eqref{eqn:D2_cases}. To understand this behaviour we first note that the second order structure function can be rewritten \cite{Pope2000} as
\begin{equation}
  D_{2}\left( r \right)=2\sigma_{\ind{u}}^{2}-2R_{\parallel}\left( r \right),
  \label{eqn:D2_R}
\end{equation}
where 
\begin{equation}
  R_{\parallel}\left( r \right)=\frac{1}{3}\bigl< \vect{u}\left( \vect{x}+\vect{r},t \right)\cdot\vect{u}\left( \vect{x},t \right)\bigr>
  \label{eqn:R_Def}
\end{equation}
is the (longitudinal) spatial velocity correlation function. The factor $1/3$ is introduced because we are only calculating the correlation of one component. At this point it is important to remark that the Hamiltonian \eqref{eqn:Hamiltonian} only correlates the directions of the fluid particles, but not their magnitudes. This can be seen by writing
\begin{equation}
  R_{\parallel}\left( r \right) = R_{\ind{a}}\left( r \right)\cdot R_{\ind{d}}\left( r \right)
  \label{eqn:R_a_R_d}
\end{equation}
with the mean of the product of the velocity magnitudes
\begin{equation}
  R_{\ind{a}}\left( r \right) = \frac{1}{3}\bigl< \lvert\vect{u}\left( \vect{x}+\vect{r} \right)\rvert\,\lvert\vect{u}\left( \vect{x} \right)\rvert\bigr>,
  \label{eqn:DefR_a}
\end{equation}
and the spatial correlation function of the directions of the fluid particle velocities
\begin{equation}
  R_{\ind{d}}\left( r \right) = \Bigl< \frac{\vect{u}\left( \vect{x}+\vect{r} \right)\cdot\vect{u}\left( \vect{x} \right)}{\lvert \vect{u}\left( \vect{x}+\vect{r} \right) \rvert\,\lvert\vect{u}\left( \vect{x} \right)\rvert} \Bigr>.
  \label{eqn:DefR_d}
\end{equation}
Fig.~\ref{fig:R_740} shows measurements of these three functions for $R_{\lambda}=740$.
From this plot it is clearly visible that the correlation function $R\left( r/\eta \right)/\sigma_{\ind{u}}^{2}$ does not approach unity in the limit of distance tending to zero, and because of that the second order structure function does not go to zero as desired. This is a result of two factors: First the directions of the fluid particles are not correlated strongly enough on small distances, as can be seen also in Fig.~\ref{fig:R_740}, since $R_{\ind{d}}\left( r/\eta \right)$ does not go to one. This behaviour is reasonable since there are only a rather small amount of fluid particles present in the system and it may not be possible to find ``a perfect match'' for every pair of fluid particles. One should note that only a small amount of particles (less than $1\%$) are located close enough to even give a contribution to the correlations on these small distances. The second reason why the correlation function $R\left( r/\eta \right)/\sigma_{\ind{u}}^{2}$ does not go to unity is that the magnitudes of the velocity vectors are not correlated. This can be seen by the fact that $R_{\ind{a}}\left( r/\eta \right)$ is almost independent on $r$. The value of $R_{\ind{a}}\left( r/\eta \right)$ can be calculated analytically:

The components of the velocity vector of fluid particles are basically normally-distributed random variables. Assuming for the time being a velocity variance $\sigma_{\ind{u}}^{2}=1\,\mathrm{m}^{2}/\mathrm{s}^{2}$ results that the absolute values of such vectors follow a $\chi$-distribution with $3$ degrees of freedom and their probability density function (PDF) is
\begin{equation}
  \chi\left( x;3 \right) = \frac{1}{\sqrt{2}\Gamma\left( \frac{3}{2} \right)}x^{2}e^{-x^{2}/2}.
  \label{eqn:pdf_chi-distribution}
\end{equation}
In general, the PDF of the product of two statistically independent random variables $x$ and $y$, with PDF $p_{\ind{X}}(x)$ and $p_{\ind{Y}}\left( y \right)$ can be calculated \cite{Springer1979} by
\begin{equation}
  p_{Z=X\cdot Y}\left( z \right) = \int_{-\infty}^{\infty}\frac{1}{\lvert x \rvert}p_{\ind{X}}\left( x \right)p_{\ind{Y}}\left( \frac{z}{x} \right)\mathrm{d}x.
  \label{eqn:pdf_Z}
\end{equation}
For two $\chi$-distributed random variables this integral gives
\begin{equation}
  p_{Z}\left( z \right) = \frac{2}{\pi}z^{2}K_{0}\left( z \right),
  \label{eqn:p_Z_chi}
\end{equation}
where $K_{0}\left( z \right)$ is a modified Bessel function of second kind. The expectation value of $z$ is then calculated to be
\begin{equation}
  \langle z \rangle=\frac{8}{\pi}.
  \label{eqn:<z>}
\end{equation}
Only considering the component parallel to $\vect{r}$ and introducing an arbitrary velocity variance $\sigma_{\ind{u}}^{2}$ gives
\begin{equation}
  R_{\ind{a}}\left( r/\eta \right) = \frac{8}{3\pi}\sigma_{\ind{u}}^{2}.
  \label{eqn:RaConst}
\end{equation}

\begin{figure}
  \includegraphics{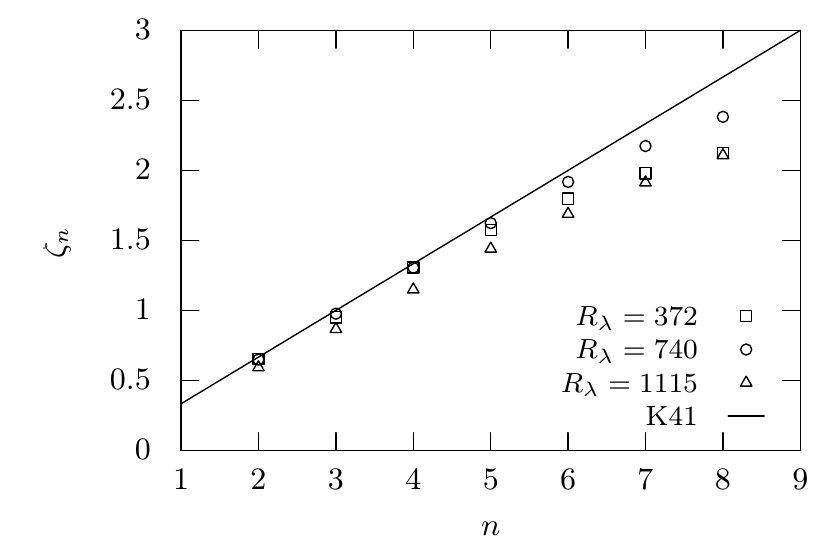}
  \caption{Numerically measured exponents of the spatial velocity structure functions up to eighth order. For $n>3$ the measured exponents are smaller than the prediction of Kolmogorov's K41 theory. This result is confirmed by experiments \cite{Anselmet1984}.}
  \label{fig:Dn_exponents}
\end{figure}

We further calculated the higher order velocity structure functions $D_{n}\left( r/\eta \right)$ up to eighth order. For $R_{\lambda}=740$ these curves are shown in the insets of Fig.~\ref{fig:D2_R}. For the other two Reynolds numbers the curves are similar. At large distances the fluid particles are not correlated anymore and the values of $D_{n}(r/\eta)$ should approach the constants given in Eq.~\eqref{eqn:Dn_Gaussian}. The resulting curves are therefore rescaled with these constants. The increase of the exponents of the power laws for intermediate distances is clearly visible. For larger values of $n$ the power laws are less pronounced and it becomes more difficult to determine the corresponding exponents. We also note, that with increasing Reynolds number the transition region from the power law to the large scale behavior becomes wider and therefore the power law region also becomes less pronounced. We measured the exponents $\zeta_{n}$ of the power laws and the results are shown in Fig.~\ref{fig:Dn_exponents}, together with the theoretical value of Kolmogorov's K41 theory \cite{Monin1975} $\zeta_{n}=\frac{n}{3}$. %
It can be seen that the resulting exponents are different for the three considered Reynolds numbers. This difference may be reduced by further fine tuning of the parameters $J_{0}$, $\alpha$, or $r_{\ind{c}}$. The measured exponents also move away from Kolmogorov's prediction for higher orders---a property well known from experiments \cite{Anselmet1984,Pope2000}.

\section{Dispersion of fluid particles}
With the model described in the last section we are able to generate a turbulent fluid velocity field that (at least on not too small scales) obeys the desired spatial statistics. We now take a closer look at the dispersion of individual fluid particles.

We start by looking at the mean squared displacement (MSD) of single particles. For times much smaller than $T_{\ind{L}}$ one expects to observe a ballistic scaling and for large $t$ normal diffusion should be observed. This means that the MSD should behave as
\begin{equation}
  \bigl< \lvert \vect{x}\left( t \right) - \vect{x}_{0} \rvert^{2} \bigr>=
  \begin{cases}
    3\sigma_{\ind{u}}^{2}\,t^{2}        & t\ll T_{\ind{L}}\\[0.8ex]
    2\sigma_{\ind{u}}^{2}T_{\ind{L}}\,t & t\gg T_{\ind{L}}
  \end{cases},
  \label{eqn:msdDef}
\end{equation}
where $\vect{x}_{0}$ is the position of the particle at time $t=0$, and the brackets $\langle\cdot\rangle$ here mean averaging over many particle tracks.

\begin{figure}
  \includegraphics{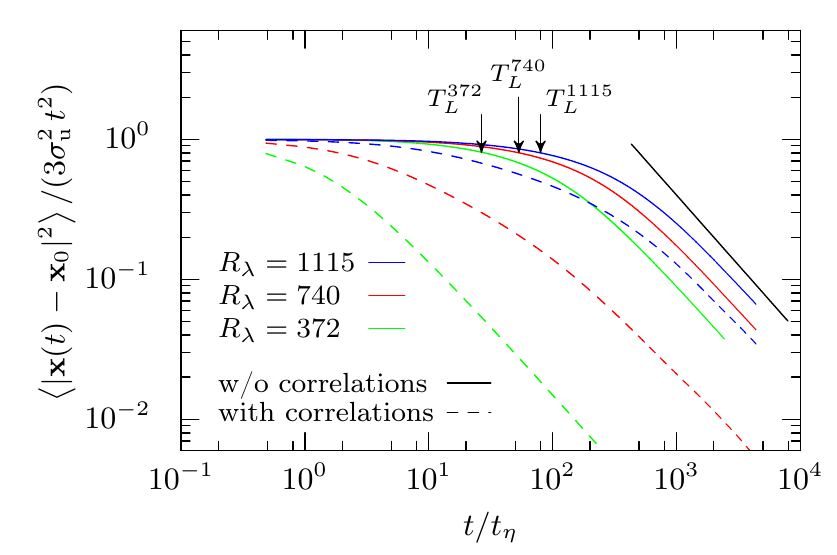}
  \caption{(Color online.) Measured mean squared displacement of fluid particles at $R_{\lambda}=372$, $740$ and $1115$ rescaled by the short time behaviour $3\sigma_{\ind{u}}^{2}t^{2}$. Simulations without correlations (solid lines) show a clear ballistic ($\sim t^{2}$) regime at short times and a transition to a normal diffusion ($\sim t$) regime for long times. The latter is indicated by the solid black line. For higher Reynolds numbers this transition happens later since $T_{\ind{L}}$ is larger for higher $R_{\lambda}$. For the cases with correlations (dashed lines) there is no clear ballistic range visible for $R_{\lambda}=372$ and $740$ and the transition to a diffusion regime happens much earlier than in the case without correlations. For $R_{\lambda}=1115$ the results are closer to the uncorrelated ones.}
  \label{fig:msd_fluid}
\end{figure}

In Fig.~\ref{fig:msd_fluid} we show the MSD rescaled by the short time behaviour for the three simulated Reynolds numbers with and without spatial correlations. %
For the simulations without correlations (corresponding to solid lines in the plot) clear ballistic scaling for small $t$ is visible. For large $t$ the transition to the normal diffusion regime (whose slope is indicated by the solid black line) is clearly observed, and since $T_{\ind{L}}$ is larger for higher $R_{\lambda}$ this transition happens later for the higher Reynolds numbers. For the simulations with spatial correlations (corresponding to dashed lines in the plot) the observed behaviour is different for the three Reynolds numbers. For $R_{\lambda}=372$ and $740$ no clear ballistic range is observed and the transition to normal diffusion is much faster than in the case without correlations. For $R_{\lambda}=1115$ on the other hand the influence of the spatial correlations is much less pronounced. For short times the $t^{2}$ scaling can still be observed and the transition to the diffusive regime happens close to the desired point at time $T_{\ind{L}}^{1115}$.

Due to the interchanging of fluid particles during correlation steps the velocity of a certain particle may change rather abruptly. This instantaneous modification of the fluid particle velocity results, on average, in a reduced Lagrangian correlation time $T_{\ind{L}}$, and therefore the transition to the normal diffusion regime happens earlier in the cases with spatial correlations. Because the Lagrangian correlation time $T_{\ind{L}}^{1115}>T_{\ind{L}}^{740}>T_{\ind{L}}^{372}$ already established spatial correlations should be preserved longer for higher Reynolds numbers. Therefore it is reasonable to assume that in a certain time interval more of the proposed fluid particle exchanges are accepted for the smaller Reynolds numbers, which again implies that the Lagrangian correlation time is reduced more for smaller Reynolds numbers than for higher ones compared to the corresponding correlation times without correlations. This may be a reason why the curve of the MSD is less influenced by the correlation steps for $R_{\lambda}=1115$.\\

Next we investigate the dispersion of pairs of fluid particles by measuring their mean squared separation (MSS). Let $\vect{\Delta}\left( t \right)$ denote the distance between two fluid particles at time $t$, and $\vect{\Delta}_{0}$ the distance of these particles at time $t=0$. For this relative dispersion one expects to observe different scaling subranges as
\begin{equation}
  \bigl< \lvert \vect{\Delta}\left( t \right) -\vect{\Delta}_{0} \rvert^{2} \bigr> =
  \begin{cases}
    \frac{11}{3}C_{2}\bigl(\langle\varepsilon\rangle\Delta_{0}\bigr)^{2/3}\,t^{2} & t_{\eta}\ll t\ll t_{0}\\[0.8ex]
    g\langle\varepsilon\rangle\,t^{3} & t_{0}\ll t \ll T_{\ind{E}}\\[0.8ex]
    4\sigma_{\ind{u}}^{2}T_{\ind{L}}\,t & t\gg T_{\ind{L}}
  \end{cases}
  \label{eqn:pairDispersionDef}
\end{equation}
where $t_{0}=(\Delta_{0}/\langle\varepsilon\rangle)^{1/3}$ and the large eddy lifetime $T_{\ind{E}}=\sigma_{\ind{u}}^{2}/\langle\varepsilon\rangle$. The first range, which is scaling with $t^{2}$, goes back to Batchelor \cite{Batchelor1950}. In this regime particles remember their initial velocity differences and therefore their motion is highly influenced by the form of the velocity structure functions $D_{n}\left( r \right)$. The second regime is known as the Richardson $t^{3}$ law, even so it was formally first introduced by Obukhov \cite{Obukhov1941} 15 years after Richardson's famous work \cite{Richardson1926} about turbulent diffusivity. The constant $g$ is known as the Richardson constant, whose exact value is still under debate \cite{Sawford2001}. The third scaling range is twice the MSD result and reflects the fact, that after a long time $t\gg T_{\ind{L}}$ the motion of fluid particles is not correlated anymore and both particles move independently from each other.

\begin{figure*}
  \includegraphics{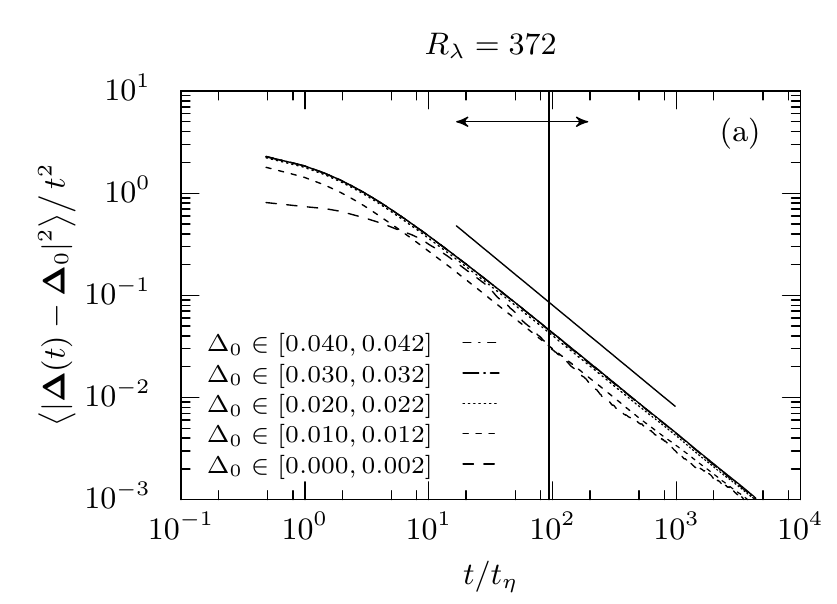}\hspace{0.5cm}
  \includegraphics{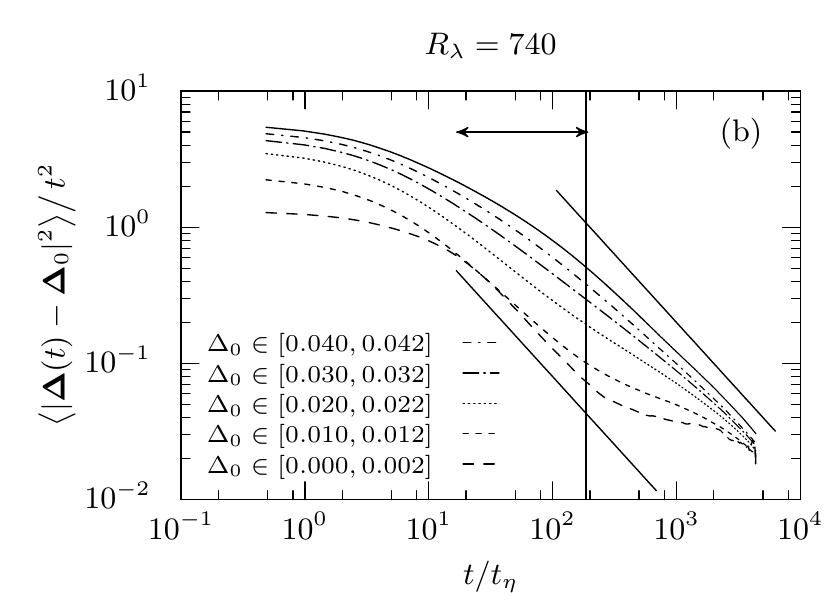}\\
  \includegraphics{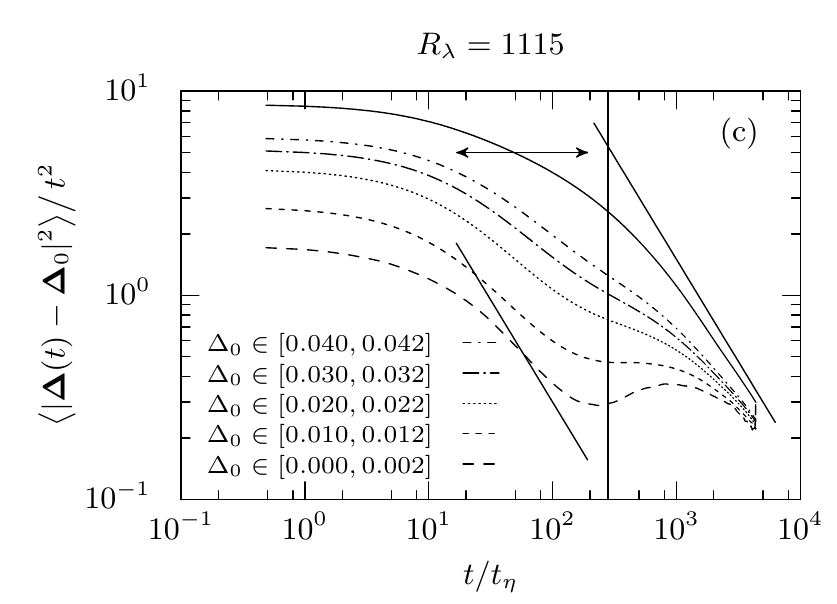}
  \caption{Measured dispersion of pairs of fluid particles rescaled by $t^{2}$. The expected Richardson $t^{3}$ could not be observed. The vertical line indicates the large eddy lifetime $T_{\ind{E}}$, the arrow shows the range of $t_{0}$ in Eq.~\eqref{eqn:pairDispersionDef}, and the transverse lines indicate normal diffusion with slope $-1$. We observe a large dependence on the initial separation between fluid particles $\Delta_{0}$ for higher $R_{\lambda}$. (a) Results for $R_{\lambda}=372$. Since $T_{\ind{E}}\sim t_{0}$ no Richardson scaling is observed. (b) Results for $R_{\lambda}=740$. For short $t$ a Batchelor $t^{2}$ scaling is observed. For $t$ in the range of $t_{0}$ the increase of the pair separation is slower than $t^{2}$. For the smallest initial separation $\Delta_{0}$ a quadratic scaling can be observed again for times larger then $T_{\ind{E}}$. (c) Results for $R_{\lambda}=1115$. Here for the smallest initial separation $\Delta_{0}$ an increase of the pair separation faster than $t^{2}$ can be found for $T_{\ind{E}}<t<1000\,t_{\eta}$, which may be an indication of Richardson scaling in this region.}
  \label{fig:pair_fluid}
\end{figure*}

Fig.~\ref{fig:pair_fluid} shows the measured relative dispersion of fluid particle pairs. %
As reported previously \cite{Sawford2008,Bourgoin2006} the behaviour of the pair separation strongly depends on the value of the initial separation $\Delta_{0}$. The solid curve corresponds to twice the measured MSD of single particles. The vertical lines indicate the time scales $T_{\ind{E}}$ at the corresponding Reynolds number. The values of $t_{0}$ range from about $17\,t_{\eta}$ at $\Delta_{0}\sim 0.001\,\mathrm{m}$ up to $t_{0}\sim 200\,t_{\eta}$ at large $\Delta_{0}\sim 0.04\,\mathrm{m}$, which are indicated by the horizontal arrow. The transverse lines have slope $-1$ to identify regions with normal diffusion.

Fig.~\ref{fig:pair_fluid}~(a) shows the resulting MSS for $R_{\lambda}=372$ rescaled with $t^{2}$. Since the eddy lifetime $T_{\ind{E}}$ is of the same order as the values of the time scale $t_{0}$ it is not surprising, that no Richardson $t^{3}$ scaling can be observed. Additionally only for short times a slight dependence on the initial separation $\Delta_{0}$ can be seen.

The results for $R_{\lambda}=740$ are shown in Fig.~\ref{fig:pair_fluid}~(b). We do not observe Richardson scaling in this case as well and therefore the data has been rescaled with $t^{2}$. For short times we observe a ballistic regime which is increasing for smaller initial separations $\Delta_{0}$. For intermediate times in the range of $t_{0}$ the pair separation becomes slower and tends to a normal diffusion for small initial separation. At the smallest initial separations $\Delta_{0}$ we observed the pair separation to almost form a plateau at $t\sim 1000\,t_{\eta}$, which would again correspond to ballistic motion. For large times all curves tend to converge to the MSD value of single particles.

In Fig.~\ref{fig:pair_fluid}~(c) we show the same measurements for $R_{\lambda}=1115$. Again a clear Richardson scaling could not be observed. This observation was also reported for experiments by Bourgoin \textit{et al.} \cite{Bourgoin2006}. The ballistic regime for short times is much more pronounced than for $R_{\lambda}=740$. For $t$ in the range of $T_{\ind{E}}$--$1000\,t_{\eta}$ we observe rather different behaviour of the pair dispersion for different values of $\Delta_{0}$. For $\Delta_{0}>0.02\,\mathrm{m}$ the pair separation is slower than in the ballistic case. For $\Delta_{0}\sim 0.01\,\mathrm{m}$ we observe a plateau, which means that the pair separation is proportional to $t^{2}$. For the smallest initial separations $\Delta_{0}<0.002\,\mathrm{m}$ we finally observe an increase in the mean distance between pairs of fluid particles which is faster than $t^{2}$. This may indicate that a Richardson scaling may be observed in this region.

The systems studied in this paper are quite dilute and particles are almost uniformly distributed inside the system at any given point in time. We therefore do not have any control over the initial separation of particles and it is rather difficult to gather enough statistics for the MSS at small $\Delta_{0}$. Thus the values of $\Delta_{0}$ were binned and the curves of the MSS in Fig.~\ref{fig:pair_fluid} were produced by averaging over all pairs within the corresponding bin. The binning and subsequent averaging may be a reason for not observing a Richardson scaling, because the number of pairs with a certain initial distance increases with $\Delta_{0}$ and therefore the averages may be biased towards the upper edge of the bin interval. This bias is particularly important for the smallest considered bin where the initial separation is most important \cite{Sawford2008} for the behaviour of the MSS.

\section{Dispersion of heavy particles}
So far we considered the motion and dispersion of fluid particles, which are by definition following the fluid streamlines perfectly. We now want to replace the fluid particles by heavy real particles and investigate the influence of the inertia of these particles on their dispersion. Therefore we introduce in addition to the $n_{\ind{p}}$ fluid particles the same number of real particles with radius $r_{\ind{p}}$, density $\rho_{\ind{p}}$ and velocity $\vect{v}_{i}$. We follow the same idea as in Ref.~\cite{Burgener2011} and ``glue'' one fluid particle to every real particle. This means that these two particles are always located at the same point in space. Energy dissipation $\varepsilon_{i}$, fluid particle acceleration $\vect{a}_{i}$ and fluid particle velocity $\vect{u}_{i}$ are still calculated by integrating Eqs.~\eqref{eqn:sde_chi}--\eqref{eqn:sde_u}, but the positions $\vect{x}_{i}$ and particle velocities $\vect{v}_{i}$ are now calculated using the discrete element model (DEM) \cite{Cundall1979}. Particles are treated as soft spheres and particle collisions are considered to be elastic with damping. Therefore we adopt the linear spring-dashpot model to resolve particle collisions. In this model particles are allowed to slightly overlap and when they do a repulsive force proportional to the overlap is applied. Further explanations of the DEM and spring-dashpot model can be found in many publications \cite{Cundall1979,Luding2008}. The (real) particles are coupled to the fluid by an empirical drag law, which gives an additional force in the DEM that accelerates the particles. For the drag force acting on particle $i$ we use the widely used and well established expression \cite{Li2003,Zhu2007,Sundaram1997,Maxey1983}
\begin{equation}
  \vect{F}_{i}^{\text{drag}} = m_{i}\frac{3}{8}\frac{C_{\ind{D}}}{r_{\ind{p}}}\left( \frac{\rho_{\ind{f}}}{\rho_{\ind{p}}} \right)\lvert \vect{u}_{i}-\vect{v}_{i} \rvert (\vect{u}_{i}-\vect{v}_{i}),
  \label{eqn:Fdrag}
\end{equation}
where $m_{i}$ is the mass of particle $i$, $\rho_{\ind{f}}$ the fluid density and the drag coefficient $C_{\ind{D}}$ is given by
\begin{equation}
  C_{\ind{D}}=\begin{cases}
    \frac{24}{\mathrm{Re}_{\ind{p}}}(1+0.15\ \mathrm{Re}_{\ind{p}}^{0.687}) & \mathrm{Re}_{\ind{p}}<1000\\
    0.44 & \mathrm{Re}_{\ind{p}}\geq 1000
  \end{cases}.
  \label{eqn:C_D}
\end{equation}
This coefficient depends on the particle Reynolds number given by
\begin{equation}
  \mathrm{Re}_{\ind{p}} = \frac{2 r_{\ind{p}}\lvert \vect{u}_{i}-\vect{v}_{i} \rvert}{\nu}.
  \label{eqn:Re_p}
\end{equation}

To quantify the influence of the inertia of particles suspended in the fluid usually the Stokes number $\mathrm{St}$ is used. It is defined by
\begin{equation}
  \mathrm{St}=\frac{\tau_{\ind{p}}}{\tau_{\ind{f}}},
  \label{eqn:St_Def}
\end{equation}
where the particle response time $\tau_{\ind{p}}$ is given by
\begin{equation}
  \tau_{\ind{p}} = \frac{4}{18}\frac{r_{\ind{p}}^{2}}{\nu}\frac{\rho_{\ind{p}}}{\rho_{\ind{f}}}
  \label{eqn:tau_p_Def}
\end{equation}
and $\tau_{\ind{f}}$ is a characteristic fluid time scale. If $\mathrm{St}$ is small the particles follow the fluid streamlines rather closely, and the larger $\mathrm{St}$ gets, the less the particle motion is influenced by the fluid. In a turbulent velocity field $\tau_{\ind{f}}$ can range from the small Kolmogorov time $\tau_{\eta}$ up to the many orders of magnitude larger energy-containing time scale $T_{\ind{L}}$. This means that a particle with response time $\tau_{\ind{p}}$ is only marginally influenced by fluid structures changing on time scales $\tau_{\ind{f}}<\tau_{\ind{p}}$.

In this paper we investigate three different particle radii $r_{\ind{p}}=10^{-5}\,\mathrm{m}$, $r_{\ind{p}}=10^{-4}\,\mathrm{m}$, and $r_{\ind{p}}=10^{-3}\,\mathrm{m}$. The particle and fluid densities are $\rho_{\ind{p}}=12.0\,\mathrm{kg}/\mathrm{m}^{3}$ and $\rho_{\ind{f}}=1.2\,\mathrm{kg}/\mathrm{m}^{3}$. This gives particle response times of $\tau_{\ind{p}}\approx t_{\eta}$, $\tau_{\ind{p}}\approx 100\,t_{\eta}$, and $\tau_{\ind{p}}\approx 10000\,t_{\eta}$. For the smallest particle size the collisions between particles were ignored, since the particle volume fraction of this system is extremely small. In this section we only present results for the two higher Reynolds numbers $R_{\lambda}=740$ and $1115$.

\begin{figure*}
  \includegraphics{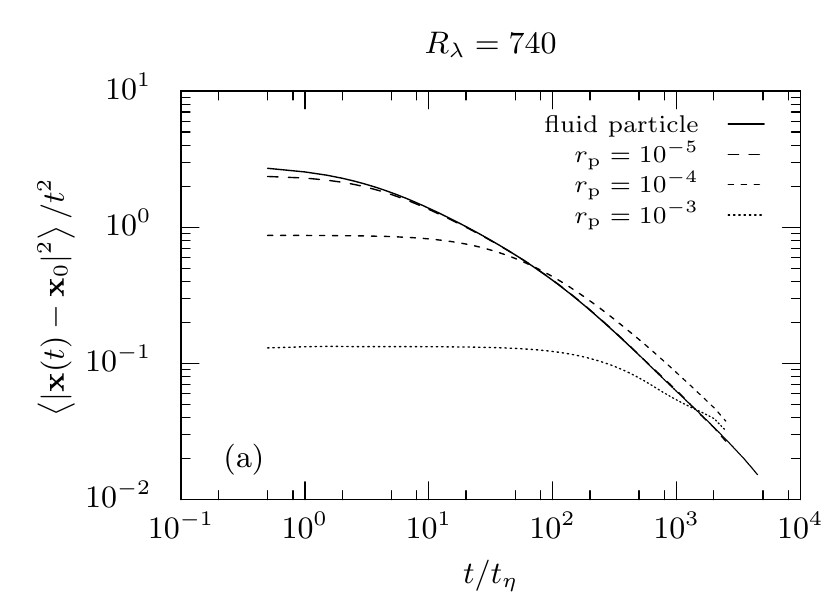}
  \includegraphics{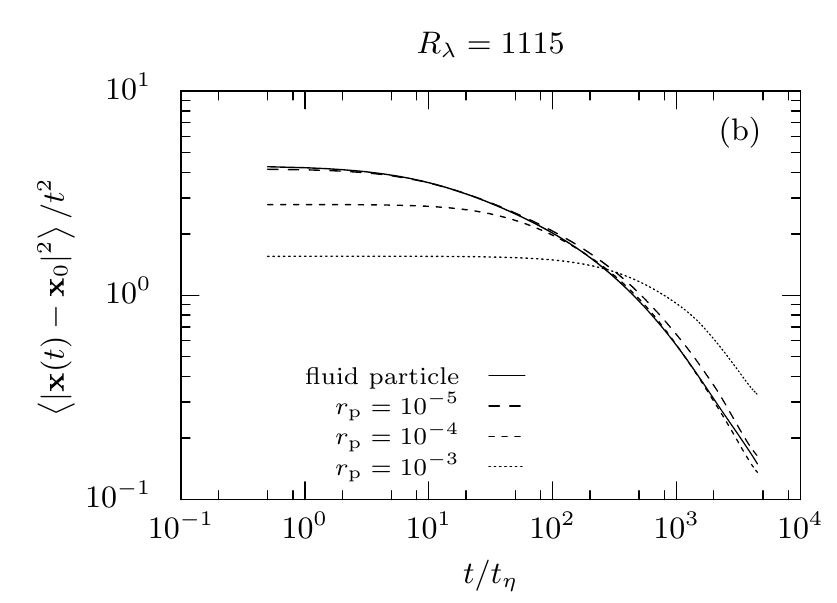}
  \caption{Numerically measured mean squared displacement for real heavy particles at two different Reynolds numbers compared to the MSD of fluid particles. The main effect of the inertia of the real particles are a reduction of the mean particle velocity and a prolongation of the ballistic range at small $t$. (a) Results for $R_{\lambda}=740$. (b) Results for $R_{\lambda}=1115$. For $r_{\ind{p}}=10^{-5}\,\mathrm{m}$ the MSD of fluid and real particles are almost the same. For $r_{\ind{p}}=10^{-4}\,\mathrm{m}$ an increase of the ballistic range at short times is visible. For long times the MSDs are again almost the same. For the largest particles at $R_{\lambda}=1115$ the ballistic range is increased further and the transition to normal diffusion happens later than for the fluid particles.}
  \label{fig:msd_wp}
\end{figure*}

Fig.~\ref{fig:msd_wp} shows the measured MSD of single particles rescaled with $t^{2}$ and compared to the results for the fluid particles. %
The short time behaviour of Eq.~\eqref{eqn:msdDef} contains the mean squared fluid particle velocity $3\sigma_{\ind{u}}^{2}$. This suggests that in Fig.~\ref{fig:msd_wp} the curves should go towards the mean squared (real) particle velocities $v_{0}^{2}\left( r_{\ind{p}} \right)$ for $t\rightarrow 0$, and indeed we can observe a mean squared velocity which is decreasing with $r_{\ind{p}}$. This is perfectly reasonable since within time intervals of length about $T_{\ind{L}}$, larger and therefore heavier particles can be accelerated on average only to smaller velocities $\vect{v}_{i}$ before the fluid velocity $\vect{u}_{i}$ changes again. Another observation we can make is that the ballistic range for small $t$ is increasing for larger $r_{\ind{p}}$, which is a result of the increasing particle response time $\tau_{\ind{p}}$.

The smallest particles with $r_{\ind{p}}=10^{-5}\,\mathrm{m}$ are of the same size as the Kolmogorov length $\eta$ and therefore one expects that the MSD of these particles is almost the same as the one of fluid particles. For larger, and therefore heavier particles their response time $\tau_{\ind{p}}$ is increased and thus the influence of the particle inertia should be visible by an enlarged $t^{2}$ range of the MSD, because the larger particles are less influenced by the fast and small scale velocity fluctuations of the fluid velocities and need more time to adjust their own velocity to the one of the fluid. Both of these effects can be seen in Fig.~\ref{fig:msd_wp} for $R_{\lambda}=740$ and $R_{\lambda}=1115$.

\begin{figure*}
  \includegraphics{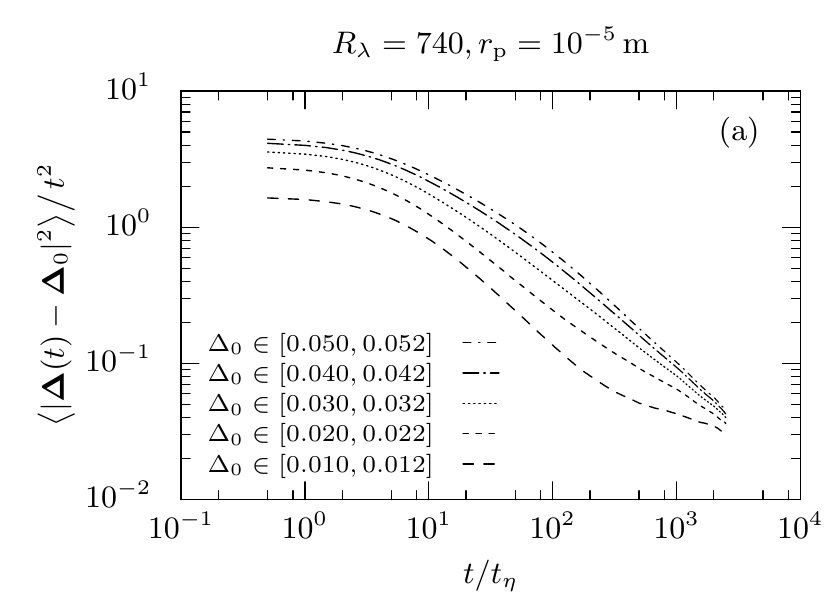}\hspace{0.5cm}
  \includegraphics{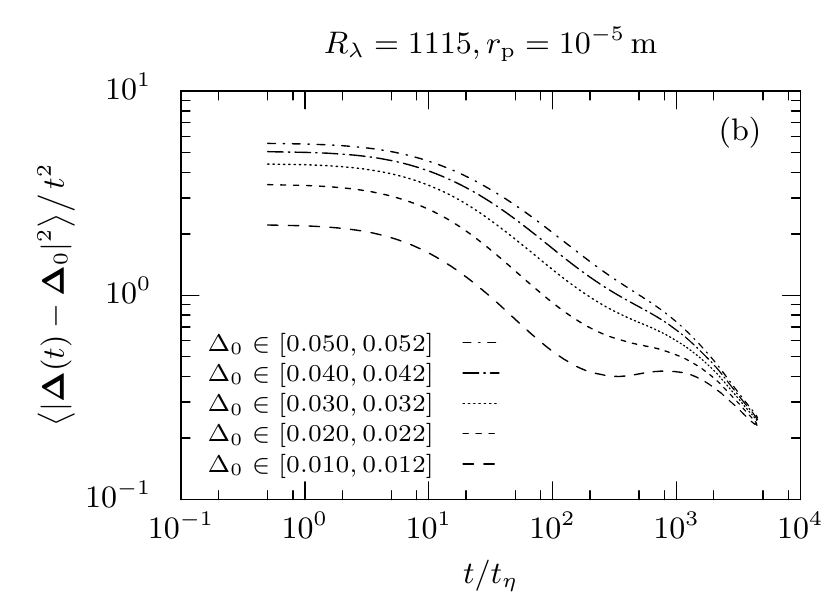}\\
  \includegraphics{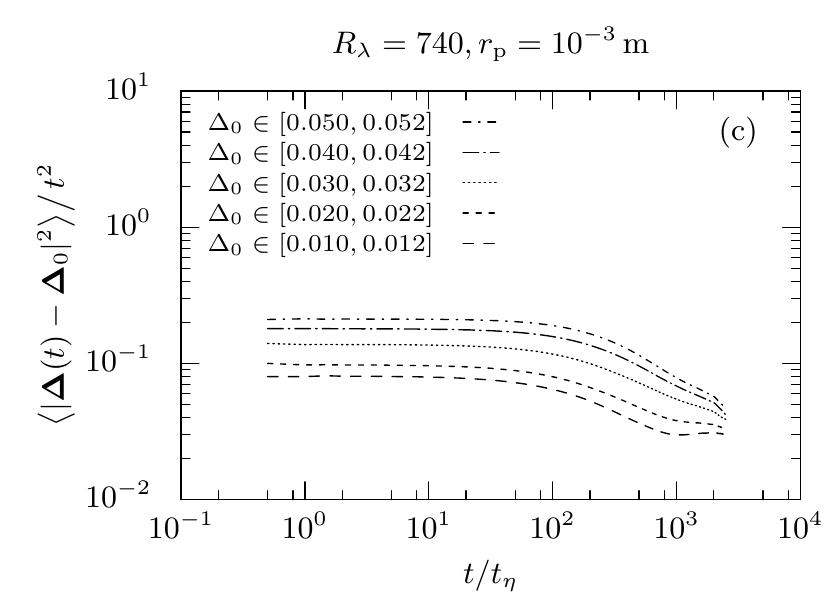}\hspace{0.5cm}
  \includegraphics{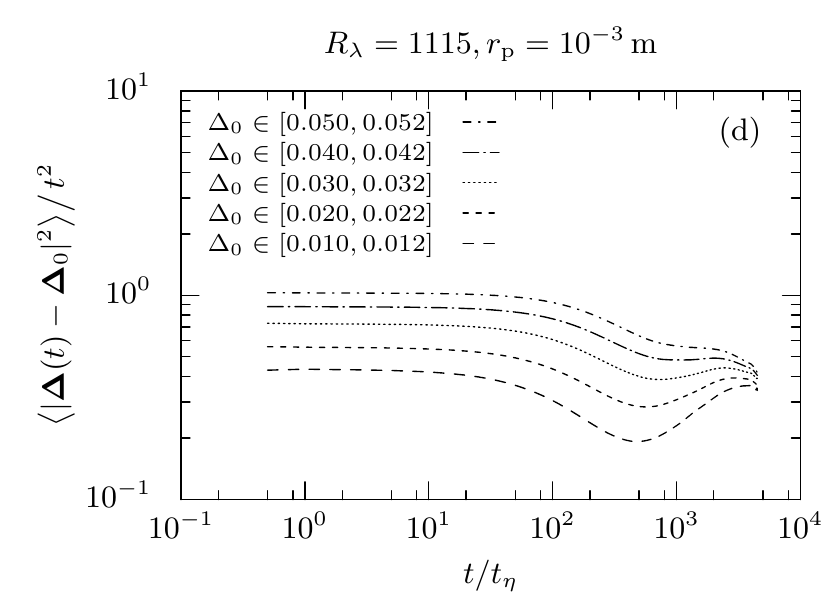}
  \caption{Numerically measured MSS for different particle radii and Reynolds numbers. (a) \& (c) Results for $R_{\lambda}=740$. (b) \& (d) Results for $R_{\lambda}=1115$. For the smallest particle radius the curves are almost the same as for fluid particles. For larger particles the $t^{2}$ regime at short times is enlarged and the mean particle velocities are reduced. For the largest particles also the long time behaviour of the MSS is influenced.}
  \label{fig:pair_wp}
\end{figure*}

Finally we again measured the dispersion of pairs of particles. The resulting MSS for two Reynolds numbers $R_{\lambda}=740$ and $R_{\lambda}=1115$ are shown in Fig.~\ref{fig:pair_wp}. %
As in the case of fluid particles we did not observe a clear Richardson scaling and therefore the results were rescaled by $t^{2}$. The main influence of the inertia of the particle can be seen by an increase of the Batchelor $t^{2}$ range at short times with increasing particle radius. Also the decrease of the mean particle velocity $v_{0}\left( r_{\ind{p}} \right)$ for larger $r_{\ind{p}}$ can be seen by a reduction of the MSS for short times compared to the case of fluid particles. For the smallest particles with radius $r_{\ind{p}}=10^{-5}\,\mathrm{m}$ (Fig.~\ref{fig:pair_wp}~(a) and (b)) the resulting MSSs are very similar to the one of fluid particles shown in Fig.~\ref{fig:pair_fluid}. As in the case of fluid particles an increase of the MSS faster than $t^{2}$ can only be observed for $R_{\lambda}=1115$. It is also interesting to note, that the Richardson regime at $R_{\lambda}=1115$ is more pronounced for larger particles. By looking at the MSD of single particles in Fig.~\ref{fig:msd_wp}~(b) one can see that for the largest particles the ballistic regime is actually larger than it is for smaller particles. This means that the integral timescale $T_{\ind{L}}$ is larger in this case, and since $T_{\ind{L}}$ and $T_{\ind{E}}$ are usually of the same order of magnitude, one can expect that the time interval in Eq.~\eqref{eqn:pairDispersionDef}, where a Richardson regime is expected, is larger for larger particles than it is for smaller particles. The prolongation of this interval then gives heavy particles ``more time'' to show a Richardson $t^{3}$ law in the MSD.

\section{Conclusion}
In this paper we presented a new method to model the dispersion of particles in turbulence. We used a set of SDEs to simulate the temporal evolution of Lagrangian tracer particles and introduced spatial correlations between them by minimizing an Heisenberg-like Hamiltonian. With this model we were able to produce turbulent velocity fields that obey the measured temporal statistics and show the correct spatial velocity structure function on distances $r>500\,\eta$. Investigations of the MSD of single particles show a ballistic regime for short times and a transition to a normal diffusion regime for large times. The algorithm for introducing spatial correlations shifts the moment in time where this transition occurs toward shorter times and this influence is smaller for larger Reynolds numbers. Further the MSS of pairs of particles were investigated. We were able to observe a Batchelor $t^{2}$ range for short times, but failed to see a clear Richardson $t^{3}$ scaling. Only for higher Reynolds numbers and small initial separations indications of a Richardson scaling could be noticed. The effects of the inertia of heavy particles on their dispersion was investigated as well. For short times an increase in the ballistic regime in the MSD as well as an increase in the Batchelor regime in the MSS could be observed with increasing particle radius. Indications of Richardson scaling were observed as well.

\begin{acknowledgments}
  The authors thank S. Succi for his useful hints and interesting discussions. We also acknowledge financial support from ETH Research Grant ETH-06 11-1.
\end{acknowledgments}

\newpage
\bibliography{references}

\end{document}